\documentclass[
    ,final            
  ]
  {aipproc}

\layoutstyle{6x9}

\begin{document}

\title{Inflaton Decay in Supergravity and Gravitino Problem}


\author{Fuminobu Takahashi}{
  address={ Deutsches Elektronen Synchrotron DESY, Notkestrasse 85,
  22603 Hamburg, Germany}
}
%


\newcommand{\gsim}{ \mathop{}_{\textstyle \sim}^{\textstyle >} }
\newcommand{\lsim}{ \mathop{}_{\textstyle \sim}^{\textstyle <} }
\newcommand{\vev}[1]{ \left\langle {#1} \right\rangle }
\newcommand{\gtrsim}{ \mathop{}_{\textstyle \sim}^{\textstyle >} }
\newcommand{\lesssim}{ \mathop{}_{\textstyle \sim}^{\textstyle <} }
\newcommand{\bear}{\begin{array}}  \newcommand{\eear}{\end{array}}
\newcommand{\bea}{\begin{eqnarray}}  \newcommand{\eea}{\end{eqnarray}}
\newcommand{\beq}{\begin{equation}}  \newcommand{\eeq}{\end{equation}}
\newcommand{\bef}{\begin{figure}}  \newcommand{\eef}{\end{figure}}
\newcommand{\bec}{\begin{center}}  \newcommand{\eec}{\end{center}}
\newcommand{\non}{\nonumber}  \newcommand{\eqn}[1]{\beq {#1}\eeq}
\newcommand{\la}{\left\langle} \newcommand{\ra}{\right\rangle}
\def\lrf#1#2{ \left(\frac{#1}{#2}\right)}
\def\lrfp#1#2#3{ \left(\frac{#1}{#2}\right)^{#3}}

\begin{abstract}
We have recently shown that, if the inflaton has a nonzero vacuum expectation value, 
it generically couples to any matter fields that appear in the superpotential at the tree level, 
and to any gauge
sectors through anomalies in the supergravity. Through these
processes, the inflaton decays into the supersymmetry
breaking sector, producing many gravitinos. The inflaton also directly
decays into a pair of the gravitinos. Taking account of these processes, 
we derive constraints on both
inflation models and supersymmetry breaking scenarios for avoiding 
overproduction of the gravitinos.
\end{abstract}

\maketitle


\section{Introduction}

Recently there has been much progress concerning the decays of scalar
fields such as moduli~\cite{moduli,Dine:2006ii,Endo:2006tf} and
inflaton~\cite{Kawasaki:2006gs,Asaka:2006bv,Endo:2006qk,Endo:2006xg,Endo:2007ih,Endo:2007sz}
in a framework of the local supersymmetry (SUSY), i.e., the
supergravity (SUGRA). The supersymmetric extension is one of the most
promising candidates for the theory beyond SM. If SUSY exists at the
TeV scale, the inflaton dynamics is quite likely described in SUGRA.
In addition, since the existence of a flat direction is mediocre in
SUSY models, one can find extremely flat potentials appropriate for
the slow-roll inflation.  Throughout this article we consider inflation
models in SUGRA. We have investigated the reheating of the universe 
in this framework, and found that the gravitinos are generically 
produced from the inflaton decay in most inflation models. In particular,
Ref.~\cite{Kawasaki:2006gs} has first pointed out that the inflaton
can directly decay into a pair of the gravitinos. Moreover,
incorporating the gravitational effects,
Refs.~\cite{Endo:2006qk,Endo:2007ih} have shown that the inflaton
generically decays into the SUSY breaking sector, which produces the
gravitinos (in)directly. The gravitino production rates due to these
processes depend on the inflaton parameters as well as the
detailed structure of the SUSY breaking sector. Such gravitino
production clearly goes beyond the simplification of the reheating
that has been adopted so far, and interestingly enough, it provides
severe constraints on inflation models as well as the SUSY breaking
scenarios. These constraints, together with the future collider
experiments and observations on CMB, should become an important 
guide to understand the high energy physics and the early universe. 
The purpose of the article is to provide both a rough sketch of the gravitino
production and the constraints on the representative inflation models.
The interested reader may refer to the original references and/or
Ref.~\cite{Endo:2007sz} for further details.

\section{Gravitino Production}

Let us first briefly review the recent development on the gravitino
production from the inflaton decay. There are three gravitino
production processes; (a) the gravitino pair
production~\cite{Kawasaki:2006gs,Asaka:2006bv,moduli,Dine:2006ii,Endo:2006tf};
(b) spontaneous decay at tree level~\cite{Endo:2006qk}; (c)
anomaly-induced decay at one-loop level~\cite{Endo:2007ih}.  For the
processes listed above, the gravitino production rate can be expressed as
\beq
\Gamma_{3/2} = \frac{x}{32 \pi} \lrfp{\la \phi \ra}{M_P}{2} \frac{m_\phi^3}{M_P^2},
\eeq
where $m_\phi$ is the inflaton mass, $\la \phi \ra$ a vacuum
expectation value (VEV) of the inflaton, and $M_P = 2.4 \times
10^{18}$GeV the reduced Planck mass.  Here it should be noted that
$\la \phi \ra$ is evaluated at the potential minimum after inflation.
The precise value of the numerical coefficient $x$ depends on the
production processes, possible non-renormalizable couplings in the
K\"ahler potential, and the detailed structure of the supersymmetry
(SUSY) breaking sector~\cite{Endo:2007sz}.  To be concrete, let us
assume the minimal K\"ahler potential and the dynamical SUSY breaking
(DSB)~\cite{Witten:1981nf} with a dynamical scale $\Lambda$. In the
DSB scenario, the SUSY breaking field $z$ can acquire a large mass
$m_z$, which is assumed to be roughly equal to the dynamical scale
$\Lambda \sim \sqrt{m_{3/2} M_P}$ in the following. Such a
simplification does not essentially change our arguments.  For a
low-inflation model with $m_\phi < \Lambda$, the process (a) becomes
effective, and $x = 1$.  On the other hand, for the inflaton mass
larger than $\Lambda$, the processes (b) and (c) become effective
instead. The inflaton decays into the hidden quarks in the SUSY
breaking sector via Yukawa couplings (process (b)), or into the hidden
gauge sector via anomalies (process (c)).  Since the hidden quarks and
gauge bosons (and gauginos) are energetic when they are produced, they
are expected to form jets and produce hidden hadrons through the
strong gauge interactions.  The gravitinos are likely generated by the
decays of the hidden hadrons as well as in the cascade decay processes
in jets.  We denote the averaged number of the gravitinos produced per
each jet as $N_{3/2}$.  Then $x$ is given
by~\cite{Endo:2007sz}~\footnote{If the K\"ahler potential takes a form
of the sequestered type, the spontaneous decay through Yukawa
couplings is suppressed~\cite{Endo:2007ih,Endo:2007sz}.}
\beq 
x \;=\; \frac{N_{3/2}}{8 \pi^2} \left(\frac{1}{2} N_y |Y_h^2| +
N_g \alpha_h^2 (T_g^{(h)} - T_r^{(h)})^2\right), 
\eeq
where $Y_h$ and $\alpha_h$ are the Yukawa coupling and a fine
structure constant of the hidden gauge group, respectively, $N_y$
denotes a number of the final states for the process (b), $N_g$ is a
number of the generators of the gauge group, and $T_g^{(h)}$ and
$T_r^{(h)}$ are the Dynkin indices of the adjoint representation and the
matter fields in the representation $r$.  Although $x$ depends on the
structure of the SUSY breaking sector, its typical magnitude is
$O(10^{-3} - 10^{-2})$ for $m_\phi > \Lambda$~\footnote{
Roughly, we expect $N_{3/2} = O(1-10^2)$, $N_g = O(1)$, $\alpha_h =
O(0.1)$, and $T_g^{(h)} - T_r^{(h)} = O(1)$, while $Y_h$ strongly
depends on the SUSY breaking models. Note also that the gravitino can
be produced through the Yukawa interaction in the messenger sector, if
the inflaton mass is larger than the messenger scale.
}.  To be explicit we take $x = 1/(8 \pi^2)$ in the following analysis. 

Using the gravitino production rate given above, we can estimate the
abundance of the gravitinos non-thermally produced by an inflaton
decay:
\bea
 Y_{3/2}^{(NT)} &=& 2 \frac{\Gamma_{3/2}}{\Gamma_{\phi}} \frac{3 T_R}{4 m_\phi},\non\\
				&\simeq & 7 \times 10^{-11}\, x \lrfp{g_*}{200}{-\frac{1}{2}} \lrfp{\la \phi \ra}{10^{15}{\rm GeV}}{2}
							\lrfp{m_\phi}{10^{12}{\rm GeV}}{2} \lrfp{T_R}{10^6{\rm GeV}}{-1},
\label{y32}							
\eea
where $g_*$ counts the relativistic degrees of freedom, and
$\Gamma_\phi \sim T_R^2/M_P$ denotes the total decay rate of the inflaton.
It should be noted that the gravitino abundance is inversely proportional to
the reheating temperature, which should be contrasted to the standard thermal
production of the gravitinos.

\section{Constraints on Inflation Models}
\label{sec:constraints-on-models}
 Now we would like to derive constraints on the inflation and SUSY
breaking models, using the non-thermal production of the gravitinos
discussed in the previous section.  The abundance of the gravitinos are tightly 
constrained by e.g. BBN,
depending on the properties of the gravitino. Using (\ref{y32}), therefore, we can 
constrain the inflaton parameters.

\begin{figure}[t]
\includegraphics[width=11cm]{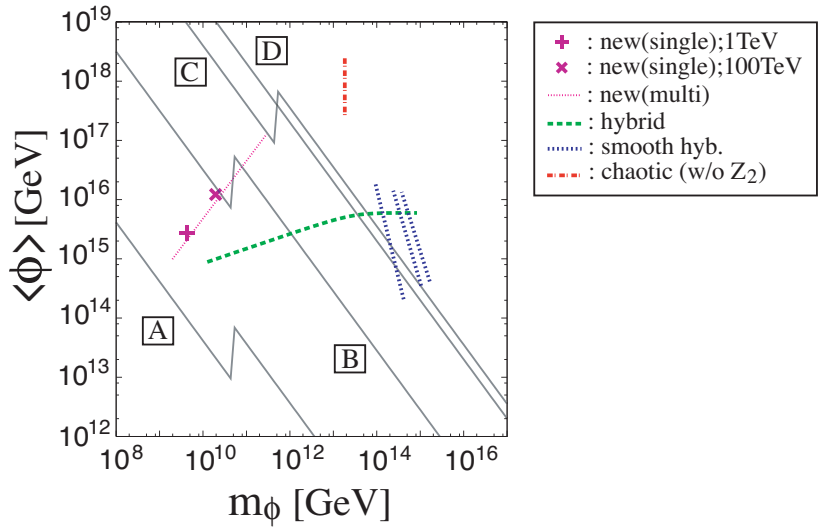}
\caption{ Constraints from the gravitino production by the inflaton
decay, for $m_{3/2} = 1{\rm\,TeV}$ with $B_h = 1$ (case\,A), $m_{3/2}
= 1{\rm\,TeV}$ with $B_h = 10^{-3}$ (case\,B), $m_{3/2} =
100{\rm\,TeV}$ (case\,C), and $m_{3/2} = 1{\rm\,GeV}$ (case\,D). The
region above the solid (gray) line is excluded for each case. 
For $m_\phi \gtrsim \Lambda$, we have used the anomaly-induced inflaton 
decay into the hidden gauge/gauginos to estimate the gravitino abundance, 
while the gravitino pair production has been used for $m_\phi \lesssim 
\Lambda$. Since $T_R$ is set to be the highest allowed value, the 
constraints shown in this figure are the most conservative ones.
}
\label{fig:m-vev}
\end{figure}

In Fig.~\ref{fig:m-vev}, we show the constraints on the inflaton mass
and VEV for $m_{3/2} = 1{\rm\,GeV}, 1{\rm\,TeV},$ and $100\,$TeV,
together with typical values of the inflation models. 
 The region above each solid line is
excluded. We find that in the case of $m_{3/2} = 1$\,TeV with the 
hadronic branching ratio $B_h=1$, all the inflation models shown in the figure are excluded. For
the gravitino mass lighter or heavier than the weak scale, the
constraints become relaxed. Indeed, if the gravitino is stable, the non-thermally produced
gravitino can account for the observed dark matter density~\cite{Takahashi:2007tz}.
The inflaton mass and its VEV depend on
the inflation models.  For larger $m_\phi$ and
$\la \phi \ra$, the constraints become severer,  because more
gravitinos are produced by the inflaton decay (see (\ref{y32})).  On the other hand, if the inflaton is
charged under some symmetries, its VEV becomes suppressed or even
forbidden especially when the symmetry is exact at the vacuum. Then
the bounds can be avoided for such inflation models. This is the case
of the chaotic inflation model with a discrete symmetry (note that the
chaotic inflation model shown in Fig.~\ref{fig:m-vev} is the one without
such a symmetry). For the other possible solutions~\cite{Endo:2006xg}, 
see Ref.~~\cite{Endo:2007sz}.

In Fig.~\ref{fig:m-vev}, we have set $T_R$ to be the highest value
allowed by the cosmological constraints. As mentioned before, the abundance of the
non-thermally produced gravitinos is inversely proportional to $T_R$,
which is different from that of the thermally produced one. 
If $T_R$ takes a smaller value, the constraints shown in the figure
becomes severer.  Thus, the bounds shown in Fig.~\ref{fig:m-vev} are
the most conservative ones. Note that one may have to introduce
couplings of the inflaton with the SM particles to realize the highest
allowed reheating temperature.

\section{Conclusion}
We have shown that the gravitinos are generically produced by
the inflaton decay, through several processes described above.
Our discovery may provide us with a breakthrough toward the full
understanding of the inflationary universe. In addition to the
standard analyses on the density fluctuations, the inflation models in
supergravity are subject to the constraints due to the (non)-thermally
produced gravitinos.  Whether a consistent thermal history after
inflation is realized now becomes a new guideline to sort out the
inflationary zoo, and hopefully it will pin down the true model,
together with data in the future collider experiments.

\bibliographystyle{aipproc}   





\end{document}